\documentclass[twocolumn,showpacs,preprintnumbers,amsmath,amssymb,superscriptaddress,prb]{revtex4}

\usepackage{graphicx,dcolumn,bm}

\newcommand{\nl}{\nonumber \\}
\newcommand{\be}{\begin{equation}}
\newcommand{\ee}{\end{equation}}
\newcommand{\bea}{\begin{eqnarray}}
\newcommand{\eea}{\end{eqnarray}}
\newcommand{\bsube}{\begin{subequations}}
\newcommand{\esube}{\end{subequations}}
\newcommand{\eq}[1]{Eq.\,(\ref{#1})}
\newcommand{\fref}[1]{Fig.\,\ref{#1}}
\newcommand{\sref}[1]{Sec.\,\ref{#1}}
\newcommand{\eref}[1]{Eq.\,(\ref{#1})}
\newcommand{\QPC}{\mbox{\tiny QPC}}
\newcommand{\rmS}{{\rm sys}}

\newcommand{\rmc}{{\rm c}}
\newcommand{\rmi}{{\rm i}}
\newcommand{\rmd}{{\rm d}}

\newcommand{\alf}{\alpha}
\newcommand{\sgm}{\sigma}
\newcommand{\Omg}{\Omega}
\newcommand{\omg}{\omega}
\newcommand{\Gam}{\Gamma}

\newcommand{\Dlt}{\Delta}

\newcommand{\dlt}{\delta}

\newcommand{\epl}{\epsilon}
\newcommand{\upa}{\uparrow}
\newcommand{\dwa}{\downarrow}
\newcommand{\GamL}{\Gamma_{\rm L}}
\newcommand{\GamR}{\Gamma_{\rm R}}
\newcommand{\gamph}{\gamma_{\rm ph}}
\newcommand{\gamd}{\gamma_{\rm d}}
\newcommand{\la}{\langle}
\newcommand{\ra}{\rangle}
\newcommand{\Tr}{{\rm Tr}}

\begin{document}

\title{Real--time counting of single electron tunneling
through a T--shaped double quantum dot system}

\author{JunYan Luo}\email{jyluo@zust.edu.cn}
\affiliation{School of Science, Zhejiang University of Science
  and Technology, Hangzhou 310023, China}
\affiliation{Department of Chemistry,
Hong Kong University of Science and Technology, Kowloon,
Hong Kong SAR, China}
\author{Shi-Kuan Wang}
\affiliation{State Key Laboratory for Superlattices and Microstructures,
 Institute of Semiconductors,
Chinese Academy of Sciences, P.O.~Box 912, Beijing 100083, China}
\author{Xiao-Ling He}
\affiliation{School of Science, Zhejiang University of Science
  and Technology, Hangzhou 310023, China}
\author{Xin-Qi Li}
\affiliation{Department of Chemistry,
Hong Kong University of Science and Technology, Kowloon,
Hong Kong SAR, China}
\affiliation{State Key Laboratory for Superlattices and Microstructures,
 Institute of Semiconductors,
Chinese Academy of Sciences, P.O.~Box 912, Beijing 100083, China}
\affiliation{Department of Physics, Beijing Normal University, Beijing 100875, China}
\author{YiJing Yan}\email{yyan@ust.hk}
\affiliation{Department of Chemistry,
Hong Kong University of Science and Technology, Kowloon,
Hong Kong SAR, China}

\date{\today}

 \begin{abstract}
 Real--time detection of single electron tunneling through a
 T--shaped double quantum dot is simulated, based on a Monte Carlo
 scheme.
 The double dot is embedded in a dissipative environment, and
 the presence of electrons on the double dot is detected with a
 nearby quantum point contact.
 We demonstrate directly the bunching behavior in electron
 transport, which leads eventually to a super--Poissonian noise.
 Particularly, in the context of full counting statistics, we
 investigate the essential difference between the dephasing mechanisms
 induced by the quantum point contact detection and the coupling
 to the external phonon bath.
 A number of intriguing noise features associated with various
 transport mechanisms are revealed.
 \end{abstract}


 \maketitle

 \section{\label{thsec1}Introduction}

 To control and manipulate electronic dynamics in nanoscale devices
 it requires  knowledge of the involving transport processes at
 single--electron level.
 The spectrum of current fluctuations, which characterizes the degree
 of correlation between charge transport events, serves as an essential
 tool superior to the average current in distinguishing various
 transport mechanisms.\cite{Bla001,Naz03}
 Full counting statistics (FCS) of current\cite{Lev964845,Bag03085316}
 has also been measured,\cite{Lu03422,Byl05361,%
 Sch042005,Van044394,Gus06076605,Fuj061634}
 owing in particular to the development of highly sensitive on--chip
 detection of single--electron tunneling technique.
 All statistical cumulants of the number of transferred particles can
 now be extracted experimentally.

 Current fluctuations would obey a Poissonian process if
 the tunneling events were statistically independent.
 However, non-Poissonian fluctuation is in general a reality.
 In the case of transport through a localized state, the Pauli
 exclusion principle suppresses the noise,\cite{Che914534}
 leading to a sub-Poissonian statistics.
 Systems of multiple nonlocal states such as coupled quantum
 dots\cite{Wie031} are more interesting. The intrinsic quantum
 coherence and many--particle interactions there result in
 different sources of correlations.
 The fascinating super-Poissionian noise thus occurs in various
 contexts, and thereby has been attracting a wide interest
 recently.

\begin{figure}
\includegraphics*[scale=0.65]{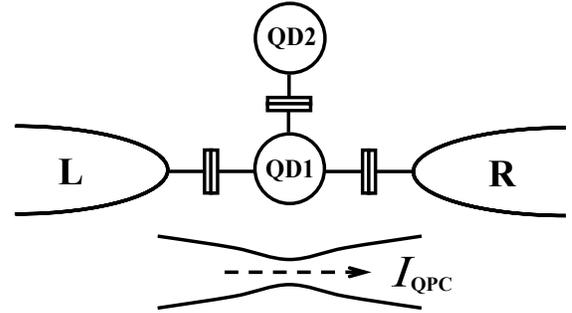}
\caption{\label{Fig1}Schematics of electron transport through a TDQD
system, monitored continuously by the QPC current that depends on
the charge state for excess electrons on two dots individually. The
TDQD is also coupled to an inevitable dissipative phonon environment
(not shown explicitly).}
\end{figure}

 A representing system, which will be studied in this work,
 is a T--shaped double quantum dot (TDQD)
 system,\cite{Kim01245326,Cor05075305,Dju05032105} as
 schematically shown in \fref{Fig1}.
 The system is of particular interest, as it can be mapped to
 a structure of quantum well in presence of an impurity inside,
 which has been investigated experimentally.\cite{Nau02161303}
 The source and drain electrodes of the TDQD are in such a
 configuration that maximizes locality versus nonlocality contrast.
 In addition, the
 TDQD is also influenced by an inevitable dissipative environment.
 The nearby quantum point contact (QPC) that serves as a charge
 detector is asymmetrically coupled to the dots.
 The current through the QPC depends on the charge state,
 ($n_1$,$n_2$), for excess electrons on two dots individually.
 Electron tunneling through the TDQD results in temporal changes
 in the charge state, and thus fluctuations in the QPC current.

 In this work, we first simulate the real--time detection of
 single electron tunneling through TDQD in experimental setup
 of \fref{Fig1}.
 The real--time simulation will be carried out based on a Monte
 Carlo approach to quantum master equation.\cite{Wan07155304}
 In particular, we demonstrate transparently the electron
 bunching, i.e., the super--Poissonian noise of charge transfer.
 The FCS is then studied, and the result shows good agreement
 with those obtained from Monte Carlo simulation. It therefore
 verifies the validity of our Monte Carlo method for
 real--time measurement.
 Particularly, it is found that the FCS is capable of differentiating
 between the dephasing effect induced by
 the QPC charge detection and that by the coupling to the
 dissipative bath environment.
 A number of intriguing noise features will be revealed in
 association with various mechanisms, demonstrating thus the
 sensitivity and selectivity of cumulants to different
 transport properties.

 The paper is organized as follows. We describe the TDQD system
 and conditional quantum master equation in \sref{thsec2}. The
 Monte Carlo simulation of real--time detection of single
 electron tunneling is summarized in \sref{thsec3}, together
 with the electron bunching behavior demonstrated vividly for
 the system in study. Having the FCS theory reviewed in
 \sref{thsec4}, we present the results and discussions in
 \sref{thsec5}. Finally, we conclude in \sref{thsec6}.

\section{\label{thsec2}Quantum master equation theory of quantum measurement}

\subsection{Model device}

 Let us start with the model setup for the QPC detection of
 single--electron tunneling through TDQD, as sketched in \fref{Fig1}.
 The total Hamiltonian consists of the coupled dots system,
 the environment, and the coupling between them; i.e.\
 $H=H_{\rm sys}+H_{\rm env}+H_{\rm sys-env}$.
 The electron Hamiltonian for the coupled dots system reads
 \be\label{Hs}
 H_\rmS=\sum_{\sgm}\big(\case{1}{2}\epl \hat Q_{z\sgm}+\Omg
 \hat Q_{x\sgm}\big)
 +\sum_l U_l \hat n_{l\upa}\hat n_{l\dwa}+U'\hat n_1 \hat n_2,
 \ee
 where $\hat Q_{z\sgm}\equiv d_{1\sgm}^\dag d_{1\sgm}-d_{2\sgm}^\dag d_{2\sgm}$,
 $\hat Q_{x\sgm}\equiv d_{1\sgm}^\dag d_{2\sgm}+d_{2\sgm}^\dag d_{1\sgm}$,
 $\hat n_l=\sum_\sgm \hat n_{l\sgm}$, and $\hat n_{l\sgm}=d_{l\sgm}^\dag d_{l\sgm}$,
 with $d_{l\sgm}$ ($d_{l\sgm}^\dag$) the electron
 annihilation (creation) operator in the QD1 ($l=1$)
 or QD2 ($l=2$).
 Each quantum dot is assumed to have only one spin--degenerate
 level ($\sgm=\;\upa$ or $\dwa$) in the bias window.
 The level detuning between the two dots is $\epl=\epl_1-\epl_2$.
 The interdot coupling strength is $\Omg$.
 The intradot and interdot Coulomb interactions, $U_l$ and $U'$,
 are both assumed to be much larger than the Fermi levels.
 We shall be interested in the double-dot Coulomb blockade
 regime,\cite{Luo07085325,Luo08345215} i.e.\ at most one
 electron resides in the TDQD.
 It can be realized in experiments by proper tuning the gate
 and bias voltages.\cite{Ono021313}

 The environment is of
 $H_{\rm env}=h_{\rm ph}+\sum_{\alpha}h_\alpha+h_{\rm QPC}$.
 It is composed of the phonon bath, the electron reservoirs
 of the source and drain ($\alpha={\rm L}$ and R) electrodes,
 as well as the QPC detector.
 Each of them is modeled as a collection of noninteracting
 particles. The phonon bath assumes
 $h_{\rm ph}=\case{1}{2}\sum_{j}\hbar\omg_j(p_j^2+x_j^2)$.
 The electron reservoirs are modeled with
 $h_{\alpha}= \sum_{k,\sgm}\epl_{\alf k}c_{\alf k\sgm}^\dag
 c_{\alf k\sgm}$
 and
 $h_\mathrm{QPC}=\sum_{p,\sgm}\epl_p c_{p\sgm}^\dag c_{p\sgm}
 +\sum_{q,\sgm}\epl_q c_{q\sgm}^\dag c_{q\sgm}$, respectively.
 These electronic parts are written in terms of electron creation
 and annihilation operators in the $\alpha$-electrode and the QPC
 reservoirs states.

 The system--environment coupling can be written as
 \bea\label{Hprm}
 H_{\rm sys-env}&=&\sum_\sgm \hat Q_{z\sgm}F_{\rm ph}
 +\sum_{\alf,\sgm}(d^\dag_{1\sgm}f_{\alf\sgm}
 +f^{\dag}_{\alf\sgm}d_{1\sgm})
 \nl
 &&+\sum_{s=0,1,2} \hat n_s F_s^{\rm QPC}.
 \eea
 The first term describes the coupling with phonon bath, in
 which $F_{\rm ph}\equiv\sum_j {\lambda}_j x_j$. This term
 is responsible for the dot level energy fluctuations. The
 effect of phonon bath on the double--dot system is
 characterized by the phonon interaction spectral density,
 $J_{\rm ph}(\omega)=\sum_j|\lambda_j|^2\delta(\omega-\omega_j)$.
 The second term describes the transfer coupling between QD1 and
 the leads, in which
 $f_{\alf\sgm}\equiv\sum_k t_{\alf k\sgm}c_{\alf k\sgm}$.
 The third term describes the interaction between the QPC detector
 and the measured system, in which
 $F_s^{\rm QPC}\equiv \sum_{p,q,\sgm} (t_{spq}c_{p\sgm}^\dag
 c_{q\sgm}+{\rm H.c.})$.
 The amplitude of electron tunneling through the QPC depends
 on the TDQD charge state $\hat n_s=|s\ra\la s|$, with
 $s=0$ for no excess electrons, and $s=1$ or 2 for one excess
 electron in QD1 or QD2, respectively.
 In other words, the current through the QPC sensitively depends
 on the charge states of the TDQD, and thus can be used to
 measure single electron tunneling events.
 The effects of these electron reservoirs components on the
 double--dot system are characterized individually by their
 interaction spectral densities:
 $J_{\alpha}(\omega)=\sum_{k\sgm}
 |t_{\alpha k\sgm}|^2\delta(\omega-\epsilon_{\alpha k})$,
 and
 $J^{\rm QPC}_s(\omega)= \sum_{p,q,\sgm} |t_{spq}|^2
 \delta(\omega - \epsilon_{p\sgm}+\epsilon_{q\sgm})$, respectively.

 In the Coulomb blockade regime, together with large source--drain
 voltage, the tunneling between QD1 and the electrode $\alf=$L or R
 can be characterized by the rate
 $\Gam_{\alpha}(\omg)=2\pi\sum_{k\sgm}|t_{\alf k\sgm}|^2
 \dlt(\omg-\epl_{\alf k})$.
 In what follows we adopt flat bands in the electrodes,
 which yields energy independent couplings $\Gam_\alf$.
 Analogously, the coupling with the QPC is described by
 the rate ${\cal T}_s=2\pi g_pg_q|t_{spq}|^2 V_{\rm QPC}$,
 where $V_{\rm QPC}$ is the QPC bias voltage, $g_{p}$
 and $g_{q}$ are the density of states in the QPC reservoirs.
 We assume hereafter the density of states to be constant, and
 $t_{spq}$ reservoir states independent.
 Thus ${\cal T}_s$ just depends on the system charge occupation
 state $|s\ra\la s|$, with $s=0$ being for zero excess electron,
 and $s=1$ or 2 for one excess electron in QD1 or QD2,
 respectively. For the QPC current, $I_s=e{\cal T}_s$, we
 have $I_0>I_2>I_1$, as implied in the scheme of \fref{Fig1}.

\subsection{Quantum master equation theory}
\label{thsec2B}

 To describe the quantum measurement, we exploit the reduced
 density operator ${\rho}^{(n)}(t)$ of the TDQD system, for the
 specified number $n$ electrons having passed through the QPC
 detector and being recorded up to the given time. The
 related conditional quantum master equation can be derived,
 which is greatly simplified under the Born--Markov
 approximation.\cite{Luo07085325,Luo08345215,Li05066803,Li05205304}
 Here, instead of using the ``$n$''--resolved equation directly,
 it is convenient to  introduce its $\chi$--space counterpart via
 $\rho(\chi,t)\equiv\sum_n e^{\rmi n\chi}\rho^{(n)}(t)$, where $\chi$
 is the so--called counting field. Let the quantum master equation
 be formally
 \be\label{CME}
 \frac{\partial}{\partial t}\rho(\chi,t)
 =-(\rmi{\cal L}+{\cal R}_\chi)\rho(\chi,t),
 \ee
 where ${\cal L}\,\cdot \equiv [H_{\rm sys},\cdot\,]$ is the
 system Liouvillian, and ${\cal R}_{\chi}$ is the dissipation
 superoperator to be specified soon. Hereafter, we set unit of
 $\hbar=e=1$ for the Planck constant and electron charge, unless
 where is needed for clarity.

 To that end, let us recast the reduced density matrix
 in the vector notation,
 \be\label{rho_vec}
 \rho=(\rho_{00},\rho_{11},\rho_{22},\rho_{12},\rho_{21},
 \rho_{\bar{1}\bar{1}},\rho_{\bar{2}\bar{2}},
 \rho_{\bar{1}\bar{2}},\rho_{\bar{2}\bar{1}})^T,
 \ee
 with $\rho_{ss'}\equiv\la s\!\!\upa\!\!|\,\rho\,|s\!\!\upa\ra$ and
 $\rho_{\bar{s}\bar{s}'}\equiv\la s\!\!\dwa\!\!|\,\rho\,|s\!\!\dwa\ra$.
 There are 9 nonzero elements of the reduced density matrix in the
 Coulomb blockade regime. The other 16 elements, describing
 coherence between different spin states, are all zeroes as the
 system--environment coupling consider here does not cause
 spin flip.
 The dissipation superoperator  ${\cal R}_\chi$ in \eq{CME} is simply
 a $9\times9$ matrix.

 In the large voltage limit and Coulomb blockade regime, the
 involving Fermi functions can be approximated by the step
 function of either one or zero.
 Under the Born-Markov approximation for weak tunnel coupling
 and second--order perturbation theory in the electron--phonon
 coupling, the dissipation superoperator ${\cal R}_\chi$ matrix
 is obtained explicitly
 \begin{widetext}
 \bea\label{calR}
 {\cal R}_{\chi}=\left(\begin{array}{ccccccccc}
 2\GamL+{\cal T}_0\,q(\chi)
 & -\GamR &  0 & 0 & 0 & -\GamR & 0 & 0 & 0 \\
  -\GamL & \GamR+{\cal T}_1\, q(\chi)
  & 0 & 0 & 0 & 0 & 0 & 0 & 0 \\
  0 & 0 & {\cal T}_2\,q(\chi)
  & 0 & 0 & 0 & 0 & 0 & 0 \\
  0 & \Gam_+ & \Gam_-  & \Gamma_\rmd & 0 & 0 & 0 & 0 & 0 \\
  0 & \Gam_+ & \Gam_-  & 0  &  \Gamma_\rmd & 0 & 0 & 0 & 0 \\
 -\GamL & 0 & 0 & 0 & 0  & \GamR+{\cal T}_1\,q(\chi)
 & 0 & 0 & 0 \\
  0 & 0 & 0 & 0 & 0 & 0  & {\cal T}_2\,q(\chi)
  & 0 & 0 \\
  0 & 0 & 0 & 0 & 0 & \Gam_+  & \Gam_- &  \Gamma_\rmd & 0 \\
  0 & 0 & 0 & 0 & 0  & \Gam_+ & \Gam_- & 0 & \Gamma_\rmd
 \end{array}\right).
 \eea
 \end{widetext}
 Here we have introduced
 \be\label{Gampm}
 \Gam_{\pm}=\frac{\pi}{2}\frac{\Omg}{\Dlt}J_{\rm ph}(\Dlt)
 \left[1\pm\frac{\epl}{\Dlt}
 \coth\Big(\frac{\Dlt}{2k_{\rm B} T}\Big)\right],
 \ee
 \be\label{Gamd}
 \Gamma_\rmd = \frac{1}{2}(\GamR+\gamph+\gamma_{\chi}),
 \ee
 and
 \be\label{gamph}
 \gamph=2\pi\frac{\Omg^2}{\Dlt^2}J_{\rm ph}(\Dlt)
 \coth\left(\frac{\Dlt}{2k_{\rm B} T}\right),
 \ee
 \be\label{gam_chi}
 \gamma_{\chi}=(\sqrt{{\cal T}_1}-\sqrt{{\cal T}_2})^2
 +2\sqrt{{\cal T}_1{\cal T}_2}q(\chi),
 \ee
 with $q(\chi)\equiv1-e^{\rmi\chi}$, $\Dlt\equiv\sqrt{\epl^2+4\Omg^2}\,$
 the Rabi frequency of the double--dot system, $k_{\rm B}$ the
 Boltzman constant, and $T$ the temperature.
 The physical processes described by the transfer elements in \eq{calR}
 are clear.\cite{Bra05315,Kie07206602,Agu04206601}

 Consider ${\cal R}^{\chi}_{11,00}=-\Gam_{\rm L}$,
 ${\cal R}^{\chi}_{00,11}=-\Gam_{\rm R}$, and their opposite spin
 counterparts, as inferred from \eq{calR}. These identities agree
 with the sequential tunneling picture, with $\Gam_{\rm L}$ being the
 rate of electron tunneling from the left electrode to system, and
 $\Gam_{\rm R}$ being that from system to the right electrode; both via QD1.

 The parameter $\Gam_{\pm}$ [\eq{Gampm}] denotes the nonsecular
 elements for the population--to--coherence transfers:
 $\Gam_+\equiv{\cal R}^{\chi}_{12,11}={\cal R}^{\chi}_{21,11}$,
 $\Gam_-\equiv{\cal R}^{\chi}_{12,22}={\cal R}^{\chi}_{21,22}$,
 and their opposite spin counterparts, as denoted in \eq{calR}.
 They are purely due to the phonon bath environment.

 The parameter $\Gamma_\rmd$ [\eq{Gamd}] is the total decoherence
 rate between two levels:
 $\Gam_\rmd\equiv {\cal R}^{\chi}_{12,12}={\cal R}^{\chi}_{21,21}$
 and the opposite spin counterparts, as denoted in \eref{calR}.
 The total decoherence rate is composed of not just $\Gamma_{\rm R}$
 due to the electron depopulation to collector, but also $\gamph$
 [\eq{gamph}] and $\gamma_{\chi}$ [\eq{gam_chi}], due to the
 phonon bath coupling and the QPC detection, respectively.
 Note that
 $\gamd\equiv\gamma_{\chi=0}=(\sqrt{{\cal T}_1}-\sqrt{{\cal T}_2})^2$
 denotes the dephasing rate induced by the existence of QPC
 ensemble.

 In the numerical demonstrations below, we set the phonon bath
 spectral density
 $J_{\rm ph}(\omega)=2\eta\omega e^{-\omega/\omega_\rmc}$. Here,
 the dimensionless parameter $\eta$ reflects the strength
 of dissipation and $\omg_\rmc$ is the Ohmic high energy cutoff.

\section{Monte Carlo simulation of single electron tunneling}
\label{thsec3}

 \begin{figure*}
 \includegraphics*[scale=1.1]{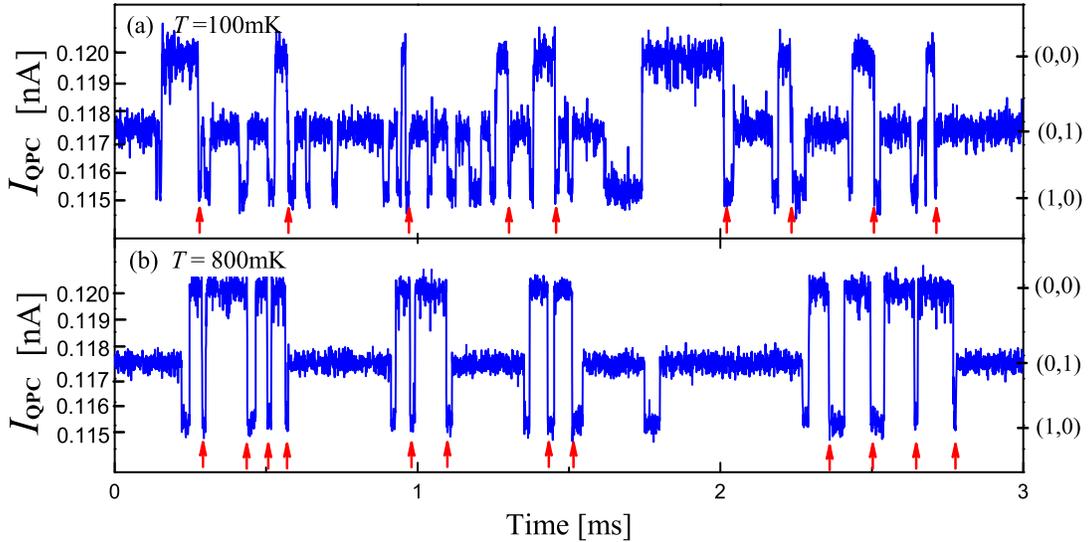}
 \caption{\label{Fig2}Time traces of the QPC current
 fluctuations correspond to different charge states
 in the TDQD (shown on the right). The arrows indicate
 transitions where an electron is entering the QD1 from
 the left lead. (a) and (b) correspond to different
 temperatures $T=100$\,mK and $T=800$\,mK, respectively.
 Other parameters are: $\GamR=4\GamL=\Omg=20$\,kHz,
 $\epl=-4\Omg$, $\eta=2\times 10^{-6}$, and
 $\omg_c=0.4$\,meV. The time step used is $\tau$=0.002ms, such that
 the minimum characteristic time-scale of the system can be
 clearly resolved.}
 \end{figure*}

 Experimentally, the most intuitive method for measuring
 the FCS of electron transport is to count electrons
 passing one by one through the conductor.
 The real--time detection of single electron transport
 enables direct evaluation of the probability distribution
 function of the number of electrons transferred through
 device within a given time period.
 In addition to the current and the shot noise, which are
 the first and second moments of this distribution, this
 method gives also access to higher order cumulants.

 Here we outline the Monte Carlo simulation scheme for
 single electron tunneling through device.\cite{Wan07155304}
 Let us start with the quantum master equation (\ref{CME}) and denote
 ${\cal L}_{\chi}\equiv -\rmi{\cal L}-{\cal R}_\chi$ there.
 The solution to the reduced state simply reads
 $\rho(\chi,t)=e^{{\cal L}_\chi\dlt t}\rho(\chi,t_0)$, for an
 arbitrary initial condition $\rho(\chi,t_0)=\rho(t_0)$ and finite
 time interval $\dlt t\equiv t-t_0$.
 Its counterpart in the particle number ``$n$''--space is
 obtained via the inverse Fourier transform
 \bea
 \rho^{(n)}(t)=\int_0^{2\pi}\!\!\frac{d\chi}{2\pi}
 e^{{\cal L}_\chi\dlt t-\rmi n\chi}\rho(t_0)\equiv{\cal U}
 (n,\dlt t)\rho(t_0).
 \eea
 The involving propagator ${\cal U}(n,\dlt t)$ is completely
 determined by the dynamic structure of the master equation (\ref{CME}),
 regardless of the initial state. Therefore, we can numerically
 evaluate it by a ``one--time task'', such as fast Fourier
 transform, which results in an efficient real--time simulation.

 Specifically,  consider the evolution of state $\rho(t_j)$ at
 $t_j$ to $\rho^{(n_j)}(t_j+\tau)$ at $t_j+\tau$:
 $\rho^{(n_j)}(t_j+\tau)={\cal U}(n_j,\tau)\rho(t_j)$,
 and denote
 ${\rm Pr}(n_j)\equiv{\rm Tr}[\rho^{(n_j)}(t_j+\tau)]$
that is the probability of having $n_j$ electrons passed
 through QPC during the time interval $[t_j,t_j+\tau]$.
 If the measurement is made but the result is ignored,
 then the (mixture) state reads
 \be\label{mixsta}
 \rho(t_j+\tau)=\sum_{n_j}\rho^{(n_j)}(t_j+\tau)
 =\sum_{n_j}{\rm  Pr}(n_j)\rho^\rmc(n_j,t_j+\tau).
 \ee
 Here,
 $\rho^\rmc(n_j,t_j+\tau)=\rho^{(n_j)}(t_j+\tau)/{\rm Pr}(n_j)$
 is the normalized state, conditioned by the definite
 number of $n_j$ electrons having passed through QPC
 during $[t_j,t_j+\tau]$.
 The second equality of \eq{mixsta} implies that if we
 stochastically generate $n_j$ according to Pr($n_j$)
 for each time interval $[t_j,t_j+\tau]$, and
 collapse the state definitely onto
 $\rho^\rmc(n_j,t_j+\tau)$, we have in fact simulated
 a particular realization for the selective state
 evolution conditioned on the specific measurement
 results.

 For the output current in a particular real--time
 measurement, we have
 \be
 I_{\QPC}(t)=I_0\rho^\rmc_{00}+I_1(\rho^\rmc_{11}
 +\rho^\rmc_{\bar{1}\bar{1}})
 +I_2(\rho^\rmc_{22}
 +\rho^\rmc_{\bar{2}\bar{2}})+\xi(t).
 \ee
 The first three terms determine the conditional evolution
 of the charge state. The last term $\xi(t)$ originates
 from the intrinsic noise of detector. Here, we consider
 in the diffusive regime, where $\xi(t)$ is a Gaussian
 variable with zero mean value and the spectral density
 $S_\xi=2e\la I_{\QPC}\ra$, with $\la I_{\QPC}\ra$ the
 average stationary QPC current.
 Accordingly, we can stochastically generate $n_j$,
 the number of electrons having passed through QPC
 during $[t_j,t_j+\tau]$, via
 $n_j =\int_{t_j}^{t_j+\tau}\!\!dt'I_{\QPC}(t') =
 [I_0\rho^\rmc_{00}+I_1(\rho^\rmc_{11}+\rho^\rmc_{\bar{1}\bar{1}})
 +I_2(\rho^\rmc_{22}+\rho^\rmc_{\bar{2}\bar{2}})]\tau
 +dW(t_j)$,
 where $dW(t_j)$ is the Wiener increment
 during [$t_j$, $t_j+\tau$].

 A typical example of simulated real--time detector current
 $I_{\QPC}$ is displayed in \fref{Fig2} for (a) $T=100$\,mK
 and (b) $T=800$\,mK, respectively.
 The temperatures are much smaller than the bias voltage
 across the TDQD, as well as the interdot and the intradot
 charging energies. Thus, electrons are transported in one
 direction, i.e.\ from the left lead to the right one.
 The corresponding charge states $(n_1,n_2)$ of the TDQD,
 with $n_1$ and $n_2$ being the excess electrons in QD1
 and QD2, respectively, are shown on the right.
 We choose rates of tunneling to the left and right leads
 as $\GamL=5$\,kHz and $\GamR=20$\,kHz, such that
 all the tunneling rates are within the bandwidth
 ($\sim30$\,kHz) of the QPC detector.\cite{Gus06076605}
 The QPC conductance (without excess electrons in TDQD) is
 $G_0=0.02\,e^2/h$, and QPC bias voltage $V_{\QPC}=0.15$\,mV,
 which corresponds to a QPC current $I_0\approx0.12$\,nA.
 The conductance is assumed to decrease by 4\% or 2\%,
 when an electron occupies on the QD1 or QD2, respectively.
 The corresponding QPC currents are $I_1\approx 0.115\,$nA
 and $I_2\approx0.118\,$nA.
 The rate of dephasing due to QPC charge detection\cite{Gur9715215}
 is then determined as
 $\gamd=\frac{1}{e}(\sqrt{I_1}-\sqrt{I_2})^2\approx 75$\,kHz.

 \begin{figure}
 \includegraphics*[scale=0.45]{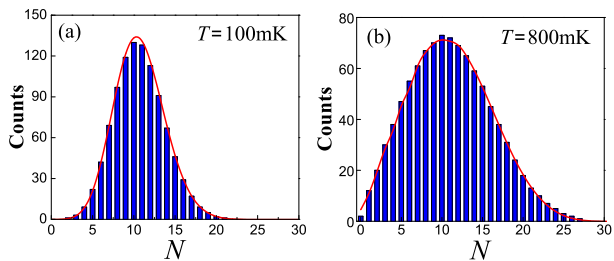}
 \caption{\label{Fig3}Statistical distribution of the
 number $N$ of electrons entering the TDQD during a
 given measurement time span of $t_\rmc=3$\,ms for
 (a) $T=100$\,mK, and (b) $T=800$\,mK, respectively.
 The other parameters are the same as in \fref{Fig2}.
 The solid line shows the distribution calculated
 from \eq{Distrib}.}
 \end{figure}

 In \fref{Fig2}, each arrow indicates the tunneling of one electron
 from the left lead into QD1. Accordingly, a charge state
 transition (0,0)$\rightarrow$(1,0) occurs.
 Remarkably, we observe a clear signature of electron bunching,
 especially when temperature is high, as shown in \fref{Fig2}(b).
 This phenomenon can be understood in terms of dynamical
 channel blockade, as will be explained later in more detail.
 For a thorough analysis of the statistical properties for
 electron transport through the TDQD device, we resort
 to the FCS. It is enabled by the probability distribution
 for the number $N$ of electrons entering the QD1 from the
 left lead (the down steps indicated by arrows in \fref{Fig2})
 during a given time span $t_\rmc$, by counting the number
 of electrons from the time traces analogous to the ones
 shown in \fref{Fig2}.
 The obtained current distributions for $T=100$\,mK and
 $T=800$\,mK are displayed by histograms in \fref{Fig3}(a) and
 (b), respectively, by repeating this counting procedure on
 one thousand independent traces with the measurement
 time span of $t_\rmc=3$\,ms.

 \section{Full counting statistics theory}
 \label{thsec4}
 We consider  hereafter the counting statistics of current
 through the TDQD device. The central quantity is then the
 probability that a given $N$ electrons have passed through
 the device during the counting measurement time span $t_\rmc$.
 This probability is related to the particle--number--resolved
 reduced density operator as
 $P(N,t_\rmc)\equiv\Tr\rho^{(N)}(t_\rmc)$, where the
 trace is over the system degrees of freedom.
 Note here ``$N$'' denotes the number of electrons transferred
 through the device, rather than the QPC detector considered
 earlier.
 From the knowledge of these probabilities one can easily
 derive not only the current and noise, but all the cumulants
 of the current distribution.
 The associated cumulant generating function (CGF)
 ${g}(\varphi)$ reads
 \be
 e^{g(\varphi)}=\sum_N P(N,t_c) e^{-\rmi N\varphi},
 \ee
 where $\varphi$ is the counting field on a specified
 TDQD lead.
 All cumulants of the current can be obtained from the
 CGF by performing derivatives with respect to the
 counting field
 \be
 \la I^k\ra=-(-\rmi\partial_\varphi)^k g(\varphi)|_{\varphi=0}.
 \ee
 The first three cumulants are related to the average current,
 the (zero-frequency) current noise, and the skewness,
 respectively.

 To evaluate the CGF, let us consider
 $\varrho(\varphi,t)\equiv\sum_N \rho^{(N)}(t)e^{\rmi N\varphi}$.
 Its equation of motion reads
 \be\label{dotvarho}
 \dot{\varrho}(\varphi,t)\equiv{\cal L}_\varphi\varrho(\varphi,t).
 \ee
 The involving generator ${\cal L}_\varphi$ is completely determined
 by the associated conditional master equation,\cite{Wan07155304}
 which is similar to \eq{CME} but with the number of electrons
 passing through the TDQD device being resolved.
 The formal solution to \eq{dotvarho} is
 $\varrho(\varphi,t)=e^{{\cal L}_\varphi t}\varrho(\varphi,0)$.
 Straightforwardly, the CGF is determined as
 $g(\varphi)=-\ln\{{\rm Tr}\varrho(\varphi,t_\rmc)\}$.
 Actually, we are most interested in the zero-frequency
 limit, i.e.\ the counting time $t_\rmc$ is much longer
 than the time of tunneling through the system.
 The CGF then simplifies
 to $g(\varphi)=-\lambda_{\rm min}(\varphi) t_\rmc$,
 where $\lambda_{\rm min}(\varphi)$ is the minimal eigenvalue
 of ${\cal L}_\varphi$ that satisfies $\lambda_{\rm min}
 (\varphi\rightarrow0)\rightarrow0$.\cite{Bag03085316,Gro06125315,%
 Fli05475,Kie06033312}
 With the knowledge of CGF, the distribution function
 can be readily obtained via\cite{Lev964845}
 \be\label{Distrib}
 P(N)=\int_0^{2\pi} \frac{d\varphi}{2\pi}
 e^{-g(\varphi)-\rmi N\varphi}.
 \ee

\section{Results and discussions}
\label{thsec5}

 In what follows, we focus our analysis on electron tunneling
 from the left lead to QD1. The relevant $N$ is then the number of
 electrons entering the dot from left lead, and $\varphi$ is the
 corresponding counting field.
 Note the same calculations apply to the counting for the right
 lead.
 The numerical results for the probability distribution are
 plotted by the solid lines in \fref{Fig3}.
 It shows a striking agreement with the histograms
 obtained by Monte Carlo simulation; thus, it also verifies
 the validity of our Monte Carlo method for real--time
 measurement.
 The two distributions in \fref{Fig3}(a) and (b) are rather
 different. The latter shows a broader and more asymmetric
 distribution than the former.  We characterize the
 differences quantitatively based on the FCS analysis,
 as follows.

\begin{figure}
 \includegraphics*[scale=0.7]{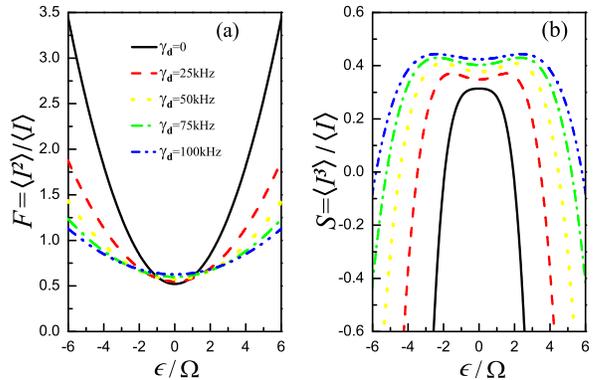}
 \caption{\label{Fig4}(a) Fano factor and (b) normalized
 skewness versus level detuning for different QPC--induced
 dephasing rates. The electron charge is set to be
 $e=1$. The temperature is $T$=100mK, and the other
 parameters are the same as in \fref{Fig2}, but in the
 absence of phonon bath; i.e., $\eta=0$. }
 \end{figure}

 Consider first the situation without electron-phonon
 interaction ($\eta=0$). The first current cumulant gives the average
 current $\la I\ra=2\GamL\GamR/\Gam_{\rm eff}$,
 with $\Gam_{\rm eff}\equiv4\GamL+\GamR$ the total effective
 tunneling width.\cite{Luo07085325,Luo08345215} It is
 independent of level detuning $\epl$, interdot coupling $\Omg$,
 and QPC charge detection induced dephasing $\gamd$.
 In contrast to the current, valuable information can be
 extracted in the second cumulant (zero-frequency shot
 noise). By expressing it in terms of the Fano factor
 $F\equiv\la I^2\ra/\la I\ra$, we readily
 obtain
 \be\label{FanoF}
 F=1-\frac{8\GamL\GamR}{\Gam_{\rm eff}^2}
 +2\GamL^2\GamR\frac{(\gamd+\GamR)^2+4\epl^2}
 {(\gamd+\GamR)\Gam_{\rm eff}^2\Omg^2}.
 \ee
 It thus allows us to get more knowledge than the current
 on the processes involved in the electronic transport.
 In \fref{Fig4}(a), the Fano factor $F$ is plotted against
 the level detuning $\epl$ for different QPC--induced dephasing rates
 $\gamd$.

 Noticeably, a significantly enhanced Fano factor is
 expected when the dot levels are far from resonance.
 In this case, electron transitions between the two dots
 are suppressed.
 For instance, if an electron is tunneled into QD2, it will
 dwell on it for a long time. In the strong Coulomb blockade
 regime, the electron in QD2 will block the current until
 it is removed and tunneled out to the right lead.
 Consequently, a mechanism of dynamical channel blockade is
 developed,\cite{Luo08345215,Kie07206602,Cot04206801,Gat02115337,%
San07146805,Kie0437,Cot04405}
 and the system exhibits strong electron bunching behavior,
 which leads eventually to a profound super--Poissonian noise.

 It is worth noting that a strong super--Poissonian noise is
 more readily achieved for a small $\Omg$ [cf.\ \eq{FanoF}].
 In this case, electron tunneling between the two dots is
 suppressed, which would enhance electron localization of
 electron in QD2. Electrons thus tend to be transferred in
 bunches.
 Remarkably, as $\Omg\rightarrow0$ a diverging Fano factor
 is found.
 Here, the strong bunching behavior is responsible for the
 divergency.\cite{Li091707}
 Similar results were also reported in Ref.\ \onlinecite{Urb09165319},
 where counting statistics for electron transport through
 a double--dot Aharonov--Bohm interferometer was investigated.
 However, in that case the divergence is closely related to
 a separation of the Hilbert space of the double--dot into
 disconnected subspaces that contain the spin singlet and
 triplet states for double occupancy.

 The QPC charge detection has a two--fold effect on the
 measured system.
 On one hand, it causes dephasing between the two dot--levels.
 Generally, the dephasing mechanism leads to the suppression of
 noise,\cite{Kie07206602} which explains the $\gamd$--dependence
 of the Fano factor shown in \fref{Fig4}, i.e.\ the noise is
 reduced with rising dephasing rate, particularly in the regime
 far from resonance.
 On the other hand, the measurement gives rise to the so--called
 quantum ``Zeno'' effect,\cite{Kor01165310,Gur03066801,Luo09385801}
 which dominates in the regime of large dephasing rate, and
 results in a strong dynamic charge blockade behavior. The noise
 is finally enhanced with increasing dephasing rate, as we have
 checked (not shown in \fref{Fig4}).

 The Fano factor studied so far proves to be much sensitive
 than the average current, it only reveals, however, limited
 information about the QPC charge detection induced dephasing
 mechanism. We thus expect more information to be extracted
 in the next order cumulant, i.e.\ the skewness.
 The numerical results for the normalized skewness
 $S\equiv\la I^3\ra/\la I\ra$ is displayed
 in \fref{Fig4}(b) as a function of level detuning $\epl$.
 Without QPC charge detection induced dephasing, the skewness
 reaches the maximum when the dot levels are in resonance
 ($\epl=0$).
 As the dephasing rate grows, it turns into a local minimum.
 At the edges of the resonance a double maximum
 structure symmetric around $\epl=0$ is observed.
 The local maxima shift away from resonance with increasing
 dephasing rates, as clearly demonstrated in \fref{Fig4}(b).

 Now let us turn to the influence of phonon heat bath that induces
 dephasing between two dots. We will reveal
 the essential difference between the dephasing induced via
 QPC charge detection and that by  phonon coupling.
 The former can be modified via the coupling between the
 TDQD and the QPC, while the latter is generated with emission
 and absorption of phonons and increases with rising temperature.
 Here, we limit our discussions to the temperatures well below
 the Coulomb charging energies and the bias voltage. Therefore
 the temperature acts solely due to the coupling to the phonon heat bath.

\begin{figure}
 \includegraphics*[scale=0.70]{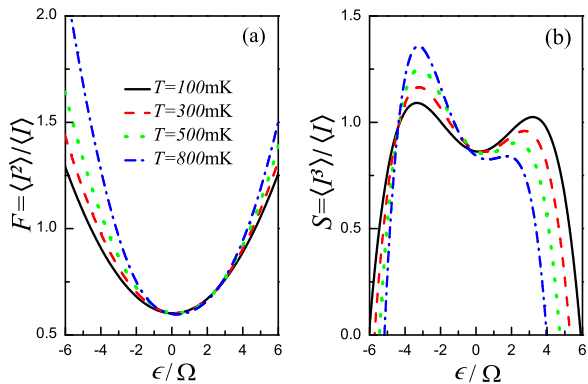}
 \caption{\label{Fig5}(a) Fano factor and (b) normalized
 skewness versus level detuning for different temperatures
 with $\gamd=75$\,kHz. The electron charge is set to be
 $e=1$. The other parameters are the same
 as in \fref{Fig2}, including those of phonon heat bath
 of $\eta=2\times 10^{-6}$ and
 $\omg_c=0.4$\,meV.}
 \end{figure}

 The calculated Fano factor and normalized skewness versus
 level detuning are plotted in \fref{Fig5} for different
 temperatures.
 Increasing temperature results in the enhancement of phonon--bath
 induced dephasing, as the emission and absorption of phonons occur
 more frequently.
 Analogous to the QPC detection, it gives rise to a mechanism
 of localization in QD2, and consequently to a dynamical
 channel blockade.
 Eventually, electron transport through QD1 occurs in
 bunches during lapses of time when the QD2 is empty.
 This is confirmed by the real--time trajectory
 [cf.\ \fref{Fig2}(b)].
Thus the noise increases with rising temperature, as displayed
 in \fref{Fig5}(a).

 If the TDQD is coupled only to the QPC, the cumulants are
 symmetric around $\epl=0$, as shown in \fref{Fig4}. However,
 with non--zero coupling to the heat bath the noise exhibits a
 clear asymmetry (see \fref{Fig5}).
 This is essentially due to another consequences of phonon
 coupling: The phonon mediated transition can partially resolve
 the dynamical charge blockade.\cite{Kie07206602,Bra091048}
 For instance, at zero temperature and $\epl>0$, spontaneous
 phonon emission can partially lift the localization caused
 by dephasing, and the spectrum turns out to be asymmetric.
 Similar argument applies to the regime of finite temperature,
 where both phonon absorption and emission take place.
 The fact that emission is more likely than  absorption explains
 the observed asymmetry.

 The skewness of the current distribution, which was not
 explored in Ref.\,\onlinecite{Kie07206602}, is found to be more
 sensitive to thermal phonon bath--induced dephasing.
 Without coupling to the phonon bath, the skewness shows a
 symmetric double maximum structure.
 The spectral becomes asymmetric in the presence of
 electron--phonon interaction.
 With increasing temperature, the phonon absorption
 takes place more frequently at $\epl<0$ , and thus greatly
 enhance the local maximum. While in the opposite regime of
 $\epl>0$ the local maximum is reduced due to suppressed
 phonon emission.

\section{Summary}
\label{thsec6}

 We have investigated electron transport through a T--shaped
 double quantum dot system by utilizing a quantum master
 equation approach.
 The major advantage of the present approach is its simplicity
 of treating properly the dephasing mechanism of the QPC charge
 detection and that due to an external heat bath.
 In addition, this approach has the merit of dealing with other
 sources of dephasing, such as that entailed by
 anti-resonances.\cite{Tor06164}
 Particularly, based on the Monte Carlo scheme, real-time detection
 of single electron tunneling is simulated by exploiting the
 sensitivity of a current, passing through a nearby quantum point
 contact, to the fluctuating charge on the quantum dots.
 Owing to the interplay between the Coulomb interactions
 and the dephasing mechanisms, a strong bunching behavior
 in the charge transfer was detected, which leads eventually
 to a super--Poissonian noise.

 Furthermore, full counting statistics of the transport current
 is analyzed based on the probability distribution,
 which is determined by ensemble
 average over a large number of single trajectories.
 It is demonstrated that the dephasing mechanism of the QPC
 charge detection and that owing to the external heat bath
 give rise to distinct and intriguing features. It thereby
 enables us to achieve a clear identification of different
 dephasing sources.
 Investigations of various processes involved in the electronic
 transport through similar devices are highly desirable in
 experiments.

 \begin{acknowledgments}
 Support from the National Natural Science Foundation of
 China under Grants No. 10904128, and the Research Grants
 Council of the Hong Kong Government (Grant No. 604709)
 are gratefully acknowledged.
 \end{acknowledgments}


\end{document}